\documentclass[11pt,twoside]{article}
\usepackage{asp2010}

\resetcounters

\begin{document}

\title{Properties of hot subdwarfs in the {\it GALEX}
survey}
\author{P. Nemeth, A. Kawka and S. Vennes}
\affil{Astronomick\'{y} \'{u}stav AV \v{C}R, CZ-25165 Ond\v{r}ejov, Czech
Republic}

\begin{abstract}
We have analyzed a sample of hot subdwarfs (sdB, sdO) selected from the
{\it GALEX} ultraviolet sky survey. Applying a model atmosphere analysis we
determined the temperature, surface gravity, and helium-to-hydrogen
abundance ratio, and obtained preliminary constraints on the CNO abundance
for a sample of 181 stars. Adopting colourimetric
(ultraviolet-infrared)
and quantitative spectral decomposition we also investigated the
incidence of solar type or earlier (A, F, G) companions.
\end{abstract}

\section{Introduction}
Hot subdwarfs are core He burning stars located at the blue end of the
horizontal branch (HB) or extreme horizontal branch (EHB). 
These subluminous objects are just below the early-type 
main-sequence (MS) stars in the Hertzsprung-Russell diagram (HRD). Being
in a relatively long lasting ($\sim$160 Myr) intermediate evolutionary 
stage of $\sim$1$M_{\odot}$ stars they are
quite common and are the primary sources of the UV excess of elliptical
galaxies, and 
overwhelm white dwarfs in blue and UV surveys of old stellar populations.

To look for bright, thus nearby white dwarf candidates in the {\it     
GALEX} database, \citet{vennes11a} devised a method
based on UV, optical and infrared colour criteria and sorted out $\sim$200
bright ($N_{\rm {\small UV}}$$<$14) and hot stars.
Between 2008 and 2011, systematic, low-resolution spectroscopic
follow-up with the ESO/NTT and NOAO/Mayall telescopes confirmed 167 stars as
hot subdwarfs. Here, we present the results of a model atmosphere
analysis of 127 sdB and 40 sdO stars. Such large and homogeneously modelled 
samples of bright subdwarfs 
are useful in identifying candidates for pulsation (e.g.,
\citealp{ostensen10})
and radial velocity studies (e.g., \citealp{geier11}).

\section{Spectral fitting}
We computed H/He/CNO non-LTE model atmospheres with {\small TLUSTY 200} and
synthetic spectra with {\small SYNSPEC 48} (\citealt{hubeny95};
\citealt{lanz95}). Model atoms of HI, HeI--II, CII--IV, NIII--V and OIV--VI; and
detailed line profiles of H and He were used in {\small SYNSPEC}. 
For all elements, we included at least three model atoms from
the {\small OSTAR} \citep{lanz03} and {\small BSTAR} \citep{lanz07}
databases. 
Model atmospheres for blue horizontal branch stars (BHB)
with effective temperatures below 20000 K were calculated with
HI, HeI--II, CII--IV, NII--IV and OII--III model atoms. 
A line list compiled from Kurucz CD-ROM 23 available at the {\small SYNSPEC} web page
was used. 

\begin{figure}
\begin{center}
 \includegraphics[width=0.4\linewidth,angle=-90]{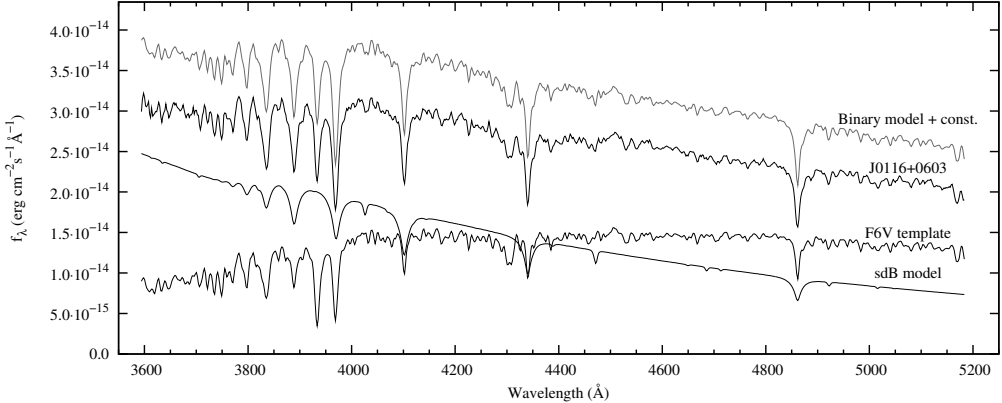}
 \caption{{\small Spectral decomposition of the sdB-F6V binary GALEX J0116+0603.}\label{Fig:0116}}
\end{center}
\end{figure}

Spectral fitting was carried out with a combination of the steepest-descent
and simplex algorithms implemented in our new $\chi^2$ minimizing
fitting program {\small XTGRID}. 
This program written in Python is an adjustable interface for {\small TLUSTY} and
{\small SYNSPEC} designed for iterative multi-wavelength spectral analysis
of hot stars from soft X-rays to near infrared wavelengths. The program
requires a starting model and with successive adjustments approaches the
observed spectrum. Quantitative 
binary spectral decomposition is also included in {\small XTGRID}. 
Figure \ref{Fig:0116} shows an example of a fit with 
binary decomposition for the sdB-F6V binary
GALEX J0116+0603.

In general, temperature and gravity 
have a higher convergence rate than abundances. To
take
advantage of this property in accelerating our procedure,
these parameters can be relaxed (kept fixed for five iterations)
after their relative changes decrease below 0.5\%. {\small XTGRID} is
scalable for cluster calculation, and to further
accelerate the fitting procedure, previously calculated models are
reused until relative changes of temperature and gravity drop below 13\% of
their respective maximum. Such accelerations are necessary to cope
with the computing demand of batch model atmosphere analyses. 

After fitting is done, parameter errors for 60, 90
and 99\% confidence intervals are estimated by mapping the $\Delta\chi^2$
with respect to the final abundance ($X$) 
at representative points in the range of $-0.99<{\delta}X/X<100$. 
Temperature and gravity errors are measured similarly in the range
of $-0.26<{\delta}T/T <0.26$ and $-0.195<{\delta}\log g/\log g <0.195$.
Parameters are
changed until the statistical limit for 60 \% confidence 
at the given number of free parameters is reached. Then, errors for 90 and 99
\% confidence are extrapolated independently 
by parabolic fits for the upper and lower error
intervals. 
In the final step, \LaTeX\ fit summaries and Gnuplot scripts are produced.

For all subdwarfs we applied the initial model with: $T_{\rm eff}=$
40000 K, 
$\log g=$ 5.6 cm/s$^2$, $\log (n$He/$n$H$)=$ -1 and $\log (n$CNO/$n$H$)=$ -2. The maximum relative
changes were limited to $\pm$5\% ($T_{\rm eff}$), $\pm$2\%
($\log g$); and $+$100\%, $-$50\% for abundances in order to maintain stable
{\small TLUSTY} convergence. 
Fitting was continued until relative changes dropped below 0.5\% for all
parameters in three consecutive iterations.
To ignore the low-order variations of the fluxed spectra, our data was sampled in 
 $80$ \AA\ sections (each having six to 16
resolution elements depending on the resolution) using the entire spectral
range.
In order to decrease model atmosphere calculation time we used 30 depth
points. Our tests showed this simplification affects model convergence 
before it would significantly change the emergent flux at low resolution.
Model atmospheres
were calculated in NLTE radiative equilibrium without convection. Detailed
profiles of H and He lines were included, but rotational broadening was
not. 
Modelling of $\sim$200 spectra took $\sim$800 hours with 6 processors (15.2
GHz total) and required the calculation of about 25000 model atmospheres.

\section{Binary decomposition}

\begin{figure}
\centering
 \includegraphics[width=0.6\linewidth]{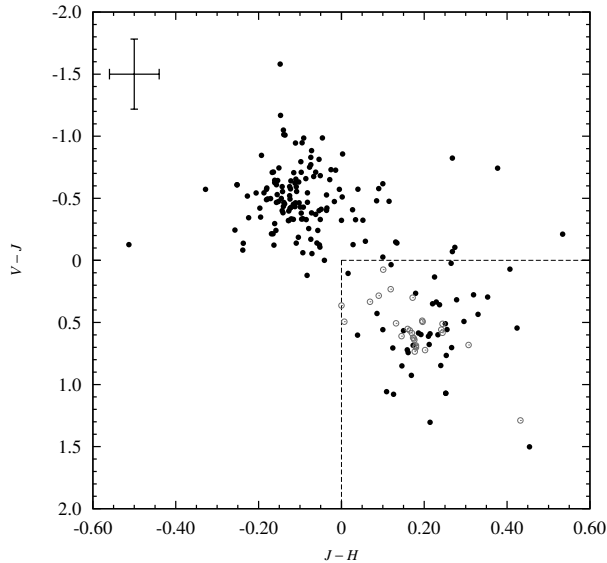}
 \caption{\small 
 $V$$-$$J$ vs. $J$$-$$H$
 colour--colour diagram for our subdwarf
 sample. Optical magnitudes were collected from GSC 2.3.2 and infrared from
 the
 2MASS database using VizieR. Apparently single stars aggregate near
 $V$$-$$J$$\approx$$-0.5$ and
 $J$$-$$H$$\approx$$-0.15$, while composite-spectra binaries show
 infrared excess (lower right corner). About 19\% of the stars have
 $V$$-$$J>0$
 and $J$$-$$H>0$. Open circles show synthetic colour indices of
 the 27 binaries resolved in this work. Typical errors on photometric colour
 indices
 are shown in the upper left corner.
\label{Fig:IRE1}}
\end{figure}

Binarity is an important aspect of the theory of subdwarf formation and evolution.
About 40\% of subdwarfs were found in binaries in the SPY sample 
\citep{napiwotzki04}. About half of the subdwarfs show
signs of binarity either 
by having composite spectra,
radial velocity variations or spectral signatures (Ca H\&K, Fe G band or MgI
lines) of a cool companion. \citet{reed04} derived a 53$\pm$6\% sdB-MS 
binary fraction from
2MASS J-H and optical B-V colours. Two stars (GALEX J0321+4727 and J2349+3844)
from our sample have already
been confirmed to be close
sdB-MS or WD binaries \citep{kawka10} and
GALEX J1717+6757 proved to be an extremely low mass white dwarf-WD binary
\citep{vennes11b} based
on radial velocity measurements.
Further
spectroscopic and photometric follow-up should reveal other similar binaries.

A
significant number of subdwarfs that were found in composite spectra
binaries have challenged model atmosphere analyses as the stellar components
need to be reliably separated. These stars were often omitted from surveys
due to complications in carrying out this task. In our sample composite spectra
of
binaries represent at least 27 objects, or 16\% of the sample. This is in accordance with
colour measurements (Figure
\ref{Fig:IRE1}), where about 19\% of stars show significant IR excess with
$V$$-$$J>0$ and $J$$-$$H>0$. 

In double-lined binaries both components can be examined simultaneously.
Although stellar parameters can be determined with larger errors in these
binaries, their analysis is an important task. We built a library of empirical
template spectra from the {\small MILES} \citep{cenarro07} database to characterize
the cool components. Altogether, 946 spectra were included in our library
from 3525 to 7500
\AA. The best
fitting secondary spectrum was searched by interpolating in temperature,
surface gravity and metallicity along with {\small TLUSTY} model parameters
for the primary. In the first step only spectral lines were considered to
select an approximate template spectrum, and from the second iteration both
the template and the flux ratio of the components were updated. 
We consider our semi-empirical approach less
ambiguous than working with synthetic spectra for both components. 

Our method worked well for binaries with late type (F, G) companions, where
components have distinct spectral features and comparable optical brightness.
We found 21 F and 4 G type MS companions. We also recovered subdwarf-AIII
and GIII binaries, but we did not find a significant fraction of A or
earlier type companions predicted by population synthesis
\citep{han03}. 
However, proper decomposition of such binaries would require
UV--optical observations to fit both components.

\section{Subdwarf atmospheric parameters}

Systematic shifts arise when data from different instruments are modelled
with various model atmosphere codes and assumptions. Our sample is large
enough and free of such major systematics, therefore appropriate to revisit the
distribution of stars in the $T_{\rm eff}$$-$$\log g$ and $T_{\rm
eff}$$-$${\rm He}$ planes and to look for possible correlations between
surface temperature, gravity and He abundance. The $T_{\rm eff}$$-$$\log g$
diagram is a very important tool in
tracking subdwarf evolution, pulsation modes and testing population
synthesis predictions. 

Out of the 181 stars, we found 127 sdB and 40 sdO stars. This number ratio
of $\sim$3.2 is close to the previously determined 3 \citep{heber09}. We found
five
He-sdB (or $\sim$4\%) among the 127 sdBs and 23 He-sdOs (or $\sim$57\%) out of the 
40 sdO stars. Of the 27 resolved binaries only four have sdO primaries.

\subsection{The $T_{\rm eff}-\log g$ plane}\label{Sec:tg}

\begin{figure}
\begin{center}
 \includegraphics[width=0.666\linewidth,angle=-90]{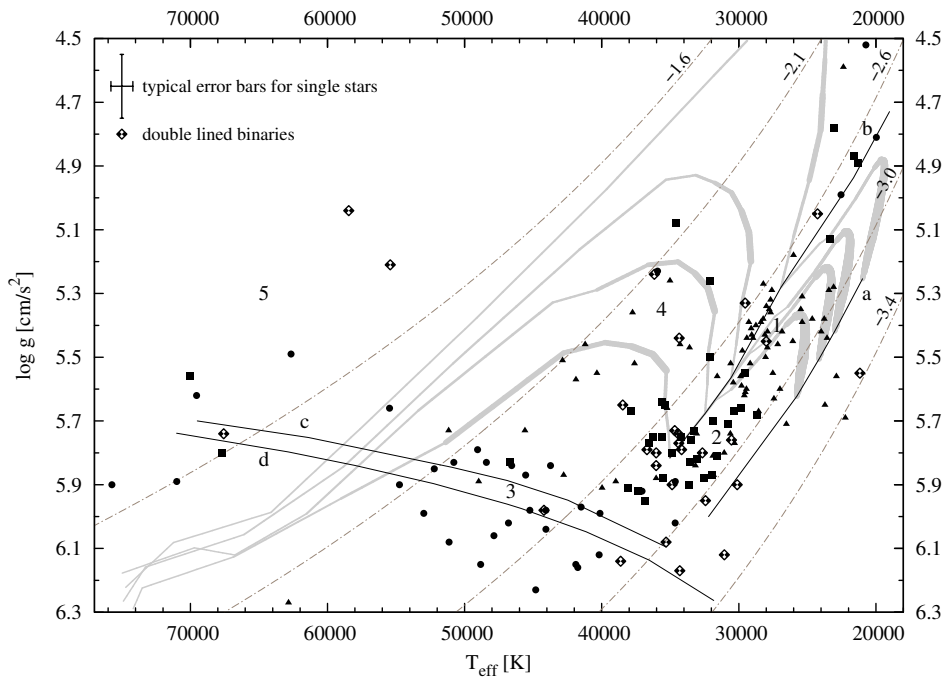}
 \caption{{\small
Sample of 181 hot stars on the $T_{\rm eff}$$-$$\log g$ diagram.
Typical error bars are shown in the upper left corner. }
\label{Fig:hrd}}
\end{center}
\end{figure}  

\begin{figure}
\begin{center}
 \includegraphics[width=0.6\linewidth,angle=-90]{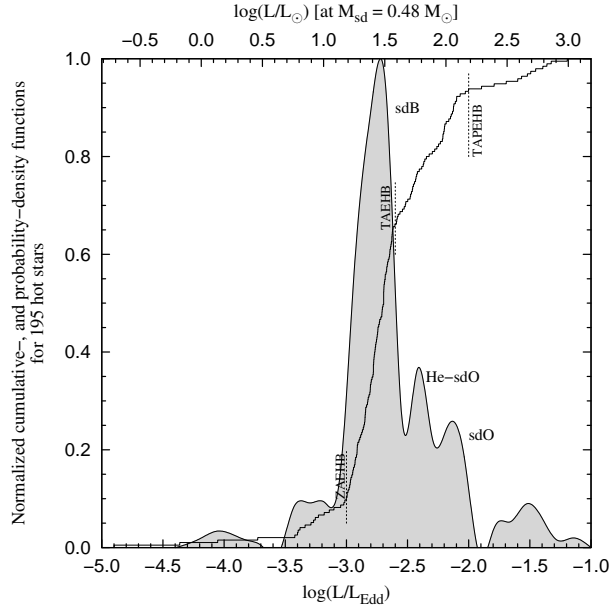}
 \caption{Luminosity distribution of 195 hot stars in the GALEX sample.
\label{Fig:lumfun}}
\end{center}
\end{figure}

Figure \ref{Fig:hrd} shows our GALEX sample in the $T_{\rm eff}-\log g$
plane. The
distribution of stars follow the theoretical EHB and He main-sequence (HeMS), with
some stars along post-EHB evolutionary tracks. 
Grey full lines show the theoretical evolutionary tracks from \citet{dorman93}
for stellar masses from top to bottom: 0.480, 0.475, 0.473 and 0.471
M$_{\odot}$. Line
widths are proportional to evolutionary time scales.
The ZAEHB and TAEHB \citep{dorman93} 
are marked with "a" and
"b", respectively. The HeMS is taken from \citet{divine65} and
\citet{paczynski71a} and are marked with "c" and "d", respectively. 
Eddington luminosity fractions are also plotted with point-dashed lines 
and marked with the
characteristic luminosities ($\log L/L_{\rm Edd}$) in the cumulative luminosity distribution
(Figure \ref{Fig:lumfun}).
He-rich $(\log (n{\rm He}/n{\rm H})>-1)$ stars are
indicated with disks, He-poor $(\log (n{\rm He}/n{\rm H})<-2.2)$
with triangles and
stars with $-2.2<\log (n{\rm He}/n{\rm H})<-1$ with squares.
Subdwarfs in
resolved composite spectra binaries are shown with diamonds.
A correlation between surface
temperature, gravity and He abundance can be immediately seen in Figure
\ref{Fig:hrd}.

We distinguish five regions, marked from 1 to 5 in Figure
\ref{Fig:hrd}. The cooler H-rich sdB stars (\#1) are found around 28000 K and $\log
g=5.45$. Many of these are possible long-period pulsators (P $\sim$ 40--170 min). The hotter sdB
stars (\#2) near 33500 K and $\log g=5.8$ are on average ten times more He
abundant and some can be short-period pulsators (P $\sim$ 2--6 min). The hot He-sdO
stars (\#3) are found over 40000 K near $\log g=6$. H-rich sdO stars (\#4) are over the
EHB and HeMS, and finally, the least defined population, possible white
dwarfs (low mass WDs, post-AGB and CSPN stars) that cross this region in the
HRD on their cooling tracks (\#5). The known shift of EHB
stars towards higher temperatures with respect to the theoretical EHB band
\citep{heber09} and the lack of stars along sections of fast evolution are notable.

From the effective temperature and surface gravity, the luminosity as a fraction of
the
Eddington luminosity can be derived assuming pure electron scattering in
fully ionized H atmospheres \citep{lisker05}. This fraction is independent
of stellar mass and proved to be a useful tool in investigating sample
completeness. Our cumulative luminosity distribution and probability
density is shown in Figure \ref{Fig:lumfun}. The change of slope in
the cumulative
luminosity function clearly marks the ZAEHB at $\log (L/L_{\rm Edd})=-3$, the
TAEHB at $\log (L/L_{\rm Edd})=-2.6$ and when stars leave the post-EHB
evolutionary phase (TAPEHB) near $\log (L/L_{\rm Edd})\sim-2.0$. A change in the slope
of sdO stars distinguishes He-rich and He-weak sdO stars. A similar overlapping of
two populations can be outlined for sdB stars as well. We found the
luminosity distribution of our sample in good agreement with previous
studies, e.g., the HS \citep{edelmann03}, the SPY sdB \citep{lisker05} and
the SPY sdO
\citep{stroeer07} surveys, and conclude that our sample is statistically complete.

\subsection{The $T_{\rm eff}-{\rm He}$ plane}

\begin{figure}
\begin{center}
 \includegraphics[width=0.900\linewidth,angle=-90]{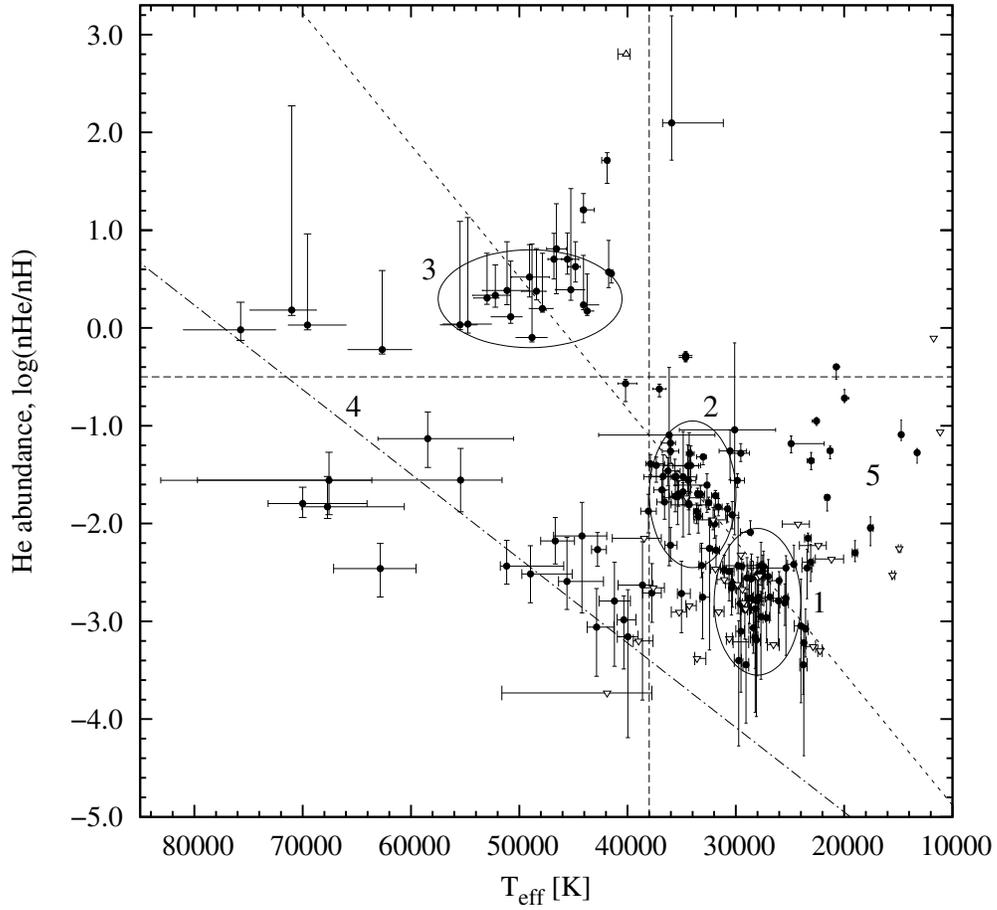}
 \caption{{\small
He abundance vs. effective temperature. Two trends showing 
 increasing
 abundance with effective temperature can be seen. Also remarkable are the  
 five marked groups introduced in Section \ref{Sec:tg}.
 Upper
 limits
 are indicated with down open triangles. }
\label{Fig:het}}
\end{center}
\end{figure}

The He abundance is another fundamental parameter of subdwarf stars that
needs to be investigated along with effective temperature and surface
gravity. Figure \ref{Fig:het} shows the He abundance vs. effective
temperature in our sample. The five groups defined in Section \ref{Sec:tg}
are well represented and marked in the figure. 
The He-rich and He-weak sequences are clearly
separated and show a similar trend with effective temperature. For the
He-rich sequence we plot the best fit line from the HS sample
\citep{edelmann03} with short-dashed line in Figure \ref{Fig:het}.
This
independently derived trend fits our data as well. However, we found a completely
different trend for the He-weak sequence, dash-dot line in Figure
\ref{Fig:het}, possibly because the temperature
range of the He-weak sequence was
under-represented in the HS survey. The lack of stars around $\log (n{\rm
He}/n{\rm H})=-0.5$ over 40000 K can be seen just like the extreme He over-abundance of
He-sdO
stars around 40000 K. One star ({\it TYC 6017-419-1}) shows a pure He
atmosphere and we could determine only a lower abundance limit. The recorded
signal-to-noise of about 100 set our detection limit near $\log (n{\rm He}/n{\rm
H})=-3.5$. 

\subsection{CN abundance}

Estimating metal abundances from optical spectra is more difficult because CNO is
relatively unimportant in calculating the model atmosphere structure and the
majority of their lines are in the UV. Out of the 181 stars, we measured a He
abundance in 148 objects, 67 showed N, and 35 showed measurable C abundance. There is
a correlation between He, C and N abundances with a number ratio of about
400:1:2.

\section{Conclusions}

We have presented a subdwarf sample of 127 sdB and 40 sdO stars from the
GALEX survey. Our selection provides a complete, homogeneously modelled
sample of hot subdwarf stars, similar to the HS and SPY surveys. This
high-quality set confirms well-known correlations between
effective temperature and He abundance. 
A dichotomy seems to emerge for sdB stars that separates stars around 28000 K
and $\log g=5.45$ from stars near 33500 K and $\log g=5.8$. This separation
is seen in the He abundance pattern and its connection to binarity 
will be investigated further. 
The full catalogue, along with further details of spectral modelling, binary
decomposition and on several individual stars will be presented in a forthcoming paper.

\acknowledgements 
We acknowledge support from the Grant Agency of the Czech Republic
(GA \v{C}R P209/10/0967) and from the Grant Agency of the Academy of Sciences of the
Czech Republic (IAA 300030908, IAA 301630901).
Observations were made with ESO telescopes at the La Silla Paranal Observatory
under programmes 82.D-0750, 83.D-0540, 85.D-0866 and with the NOAO Mayall
telescope at the Kitt Peak National Observatory.

\end{document}